# Fragmentation in a centrally condensed protostar [*]

Andreas Burkert[1] and Peter Bodenheimer[2]
[1] *Max-Planck-Institut für Astronomie, D-69117 Heidelberg, Germany*
[2] *University of California Observatories/Lick Observatory, Board of Studies in Astronomy and Astrophysics, University of California, Santa Cruz, CA 95064, USA*

**ABSTRACT**
Hydrodynamical calculations in three space dimensions of the collapse of an isothermal, centrally condensed, rotating 1 M$_\odot$ protostellar cloud are presented. A numerical algorithm involving nested subgrids is used to resolve the region where fragmentation occurs in the central part of the protostar. A previous calculation by Boss, which produced a hierarchical multiple system, is evolved further, at comparable numerical resolution, and the end result is a binary, with more than half of the mass of the original cloud, whose orbital separation increases with time as a result of accretion of high-angular momentum material and as a result of merging with fragments that have formed farther out. Repeating the calculation with significantly higher resolution, we find that a sequence of binaries can be induced by fragmentation of circumbinary disks. The stability of the resulting multiple system is investigated using $n$-body calculations, which indicate that such a system would transform on a short time scale into a more stable hierarchical structure. The outermost and most massive binary which forms in the high-resolution run has properties similar to that of the binary found in the low-resolution calculation. Thus the basic outcome is shown to be independent of the numerical spatial resolution. The high-resolution run, in addition, leads to the formation of a system of smaller fragments, which might be important for the understanding of the origin of close binaries with low-mass components and of low mass single stars.

# 1 PREVIOUS FRAGMENTATION STUDIES

Recent observational studies of the properties of pre-main-sequence binary systems (reviewed by Mathieu 1994; see also Brandner et al 1995) have confirmed that their frequency is at least comparable to that of F and G main-sequence stars in the solar vicinity. There may in fact be an excess of pre-main-sequence binaries in the Taurus-Aurigae star formation region (Ghez, Neugebauer, & Matthews 1993; Simon et al 1995). In addition, the distributions of pre-main-sequence binaries according to orbital period and according to eccentricity are very similar to those of main sequence binaries. The discovery that binary systems exist even among the very youngest T Tauri stars and among protostellar candidates (Walker, Carlstrom, & Bieging 1993; Wootten 1989) strongly suggests that the formation epoch of most systems is during the star formation and protostar collapse phases. On the theoretical front, there has been a recent explosion of numerical three-dimensional hydrodynamical calculations regarding fragmentation, either at the molecular core stage (Pringle 1989; Chapman et al 1992), during protostar collapse (Boss 1993a, Miyama 1992, Bonnell & Bastien 1993; Myhill & Kaula 1992, Sigalotti & Klapp 1994; Nelson & Papaloizou 1993; Burkert & Bodenheimer 1993, referred to as **BB**; Monaghan 1994), or after the formation of a protostellar disk (Adams, Ruden, & Shu 1989; Shu et al 1990; Adams & Benz 1992; Bonnell 1994). Although progress toward the general goals of understanding the binary frequency and the distributions of periods, eccentricities, and mass ratios has been made, the initial parameter space is so large that it is difficult to tell which initial conditions are the appropriate ones. Furthermore, the basic question of the origin of close binaries still remains to be clarified. The general mechanisms for binary formation have been recently reviewed by Pringle (1991), Boss (1993b), Bodenheimer (1992), Zinnecker (1990), and Bodenheimer, Ruzmaikina, & Mathieu (1993). A more specific review covering numerical calculations of fragmentation during protostar collapse is given by Bodenheimer (1994); a general review of angular momentum effects in protostar collapse is given by Bodenheimer (1995). While fragmentation is not the only mechanism that can provide binary systems during star formation, it is likely to be involved, at least in an initial phase. Capture through disks (Clarke & Pringle 1991a,b; Heller 1993; Ostriker 1994) is a possibility as long as the protostars have close enough separations to allow a reasonable capture rate – an initial fragmentation stage could provide the appropriate condition (see, for example, Monaghan & Lattanzio 1991). However there are difficulties with the capture process; for example high-velocity encounters can dissipate a disk (Clarke & Pringle 1991a) and relatively distant encounters can actually result in a gain of energy by the capture candidate rather than a loss (Ostriker 1994).

The problem of initial conditions can to some extent be constrained by observations. Although many of the collapse calculations have started from an initially uniform density, it is clear from observations of molecular cloud cores (Boss 1993a; Ward-Thompson et al 1994) that such structures are actually centrally condensed. Also, theoretical studies (Lizano & Shu 1989; Tomisaka, Ikeuchi, & Nakamura 1990; Ciolek & Mouschovias 1994; Basu & Mouschovias 1994) of the quasistatic evolution of magnetically supported molecular clouds end up with a centrally condensed distribution at the onset of collapse. Accordingly Boss (1991, 1993a, 1995) has assumed in his calculations that the initial density distribution is exponentially falling from the center, and that the ratio of the densities at the center and edge of the protostar is 20. This distribution is not in disagreement with observations, and it does not have the extreme central condensation of the singular isothermal sphere ($\rho \propto r^{-2}$), which, if uniformly rotating, is apparently stable against fragmentation (Myhill & Kaula 1992; Tsai & Bertschinger 1989).

In the present paper we consider the collapse of isothermal, rotating clouds, without viscosity or magnetic fields, starting with the density distribution of the type described by Boss (1991) and with uniform rotation. The initial density distribution is thus spherically symmetric with a small nonaxisymmetric perturbation. An interesting result was obtained in the Boss calculation: the cloud collapses to a ring and subsequently fragments into a hierarchical multiple with four components. The purpose of our calculation is to compare results with those of Boss (1991) at comparable numerical resolution, to carry the evolution of the fragments for a longer time than did Boss, and to investigate the effect of using a higher degree of spatial resolution than that of Boss. By following the evolution beyond the time of the initial fragmentation, up to the point where a significant fraction of the cloud mass has accumulated onto the fragments, we can determine to what extent mergers between fragments occur and what the final orbits of the resulting multiple systems are.

## 2 THE COMPUTATIONAL METHOD AND INITIAL CONDITIONS

The numerical method, which involves the use of a number of subgrids of increasingly higher resolution, is described by **BB**. The calculations are performed on a 3-dimensional Eulerian, Cartesian grid; the advection scheme is based on second-order monotonic transport (van Leer 1977). The full computational region is represented by a standard grid, composed of $64^3$ grid cells equally spaced in all directions. For improved resolution of the inner regions, up to 6 Cartesian nested subgrids can be superimposed on the standard grid. In a typical case the linear scale on a given subgrid is reduced by a factor of 2 with respect to the next larger grid. The grid structure is set up at the beginning of the calculation, and no adaptive grid procedures are used. Each grid is treated relatively independently; that is, the equations of hydrodynamics and the Poisson equation are solved throughout an entire grid, even in the regions where it overlaps with its subgrids. The grids are coupled as follows: (1) outer boundary conditions for density, gravitational potential, and velocity are obtained on each subgrid from the next-largest grid, and (2) after each timestep on an embedded grid, conserved quantities (density and momentum) are updated on the next-largest grid in the overlap region by suitable averaging. No outflow or inflow is allowed through the outer edge of the standard grid.

To allow calculation over a longer time than the approximately 1.3 initial free-fall times reported in earlier calculations (e. g. BB), a small modification of the technique is required. The previous calculations had to be stopped because of rapid uncontrolled increase in the density at the position of the dense fragments. In the present calculation, we introduce an artificial viscosity of the type described by Colella & Woodward (1984). A dissipation term, linear in the velocity gradient, is added to the momentum equation and to the continuity equation. The dimensionless viscosity parameter is set to 0.1. With this technique, the code reproduced the bar fragmentation reported by BB except that unlimited density growth in the fragments was prevented. Other techniques to deal with this problem, within the context of SPH, have been reported by Bonnell & Bate (1994) and by Bate, Bonnell, & Price (1995).

An extensive set of test calculations has been performed in order to determine the accuracy of the numerical scheme and ensure continuity of the flow across the grid boundaries; these tests are described in **BB**. For example the standard test case for fragmenting collapsing protostars (Boss & Bodenheimer 1979; Bodenheimer & Boss 1981) was followed for 1.3 free fall times and showed excellent agreement with the results of Myhill & Boss (1993).

The protostar in the present calculation is assumed to be isothermal in space and time. The initial conditions are specified by the ratios $\alpha$ and $\beta$, which are, respectively, the thermal and rotational energies divided by the absolute value of the gravitational energy, by the angular momentum distribution, by the size and form of the initial perturbation, and by the form of the density distribution. The latter is taken to be

$$\rho(r) = \rho_0 \exp\left(-r^2/r_0^2\right) \qquad (1)$$

where $r$ is the distance to the origin and $\rho_0$ and $r_0$ are constants. The form of the perturbed profile is

$$\rho_p(r) = \rho(r)[1 + a_1 \cos(2\phi)] \qquad (2)$$

where $a_1$ is the amplitude of the perturbation and $\phi$ is the azimuthal angle about the rotation ($z$) axis. In all cases presented here the density distribution is given by equations (1) and (2), $a_1$ is set to 0.1, the cloud mass is set to 1.0 $M_\odot$, $\rho_0 = 1.7 \times 10^{-17}$ g cm$^{-3}$, $r_0 = 2.9 \times 10^{16}$ cm, and the radius of the original sphere is $5 \times 10^{16}$ cm. The ratio of the central density to the density at the outer edge is 20. The parameter $\alpha$ is set to 0.26 and $\beta$ to 0.16 ($\Omega = 1.0 \times 10^{-12}$ s$^{-1}$). As the cloud is assumed to start from rigid rotation, the value of $\Omega$ completely specifies the velocity field. The sound speed $c_s = 1.9 \times 10^4$ cm s$^{-1}$. Thus in all cases these conditions are identical to those used by Boss (1991) in his case C4.

The calculations reported here (but not those of BB) assume point symmetry with respect to the $z$-axis. After each time step on each grid at each value of $z$ the density at $(x, y)$ is averaged with that at $(-x, -y)$ and both quantities are replaced by the average. The same procedure is applied to the momenta. This procedure has several advantages: (1) the calculation is more readily reproducible with an independent code, since the extreme sensitivity to initial conditions or to small fluctuations is to some extent suppressed; (2) it allows a more accurate comparison of results which use different numerical resolution, because if point symmetry were not assumed the symmetry would likely break at different times with different resolutions, and (3) the averaging procedure tends to suppress small nonsymmetric numerical fluctuations which could lead to artificial fragmentation. Note also that the Boss calculation with which we are comparing retained point symmetry throughout, although it was not specifically assumed.

## 3 RESULTS

3.1 Initial binary formation: moderate resolution

The cloud collapse, with an initial central free-fall time of 5.09 $\times 10^{11}$ s, was first calculated with three subgrids, which had radii of 1.25 $\times 10^{16}$ cm, 6.25 $\times 10^{15}$ cm, and 3.125 $\times 10^{15}$ cm. The resolution on the finest grid is 1 $\times 10^{14}$ cm, compared to 2.67 $\times 10^{14}$ cm for Boss' (1991) calculation from the same initial condition. In his case a ring-shaped off-center density maximum forms which fragments into two pairs of binaries, after 7.27 $\times 10^{11}$ s. In our case, after 7.49 $\times 10^{11}$ s the central regions of the configuration have collapsed to a thin disk with half-thickness (in $z$) of only 3 $\times 10^{14}$ cm. In the disk, a small central density maximum forms and around it a distorted ring with trailing spiral arms (Fig. 1). The ring radius is about 1 $\times 10^{15}$ cm, in good agreement with that obtained by Boss. A sequence of frames which follows the evolution of the ring into binary fragments is shown in Fig. 2. At a time of 7.543 $\times 10^{11}$ s four fragments appear in the ring (Fig. 2a), which a little later organize themselves into pairs (Fig. 2b), which are reminiscient of those found by Boss (1991; his Fig. 1d). At that point Boss stopped his calculation. As Boss notes himself, it still needs to be shown whether this configuration survives. Therefore we continue the run. As can clearly be seen from Fig. 2c to 2f, this configuration is not stable. The two leading fragments of the pair accrete material with lower angular momentum than do the trailing fragments and are pulled inwards. At $t = 7.748 \times 10^{11}$ s they begin to merge with the central density maximum (Fig. 2d). At this stage the four fragments are connected by a structure which resembles an open spiral rather than a ring. In the later evolution much of the spiral arm material winds up (Fig. 2e) and then forms a high-density disk around the central object with a diameter similar to the orbital separation of the outer binary (Fig. 2f). As a consequence of the binary formation out of a ring, its initial orbit is nearly circular and stable, with a mean separation of 4 $\times 10^{15}$ cm and a period of 1.4 $\times 10^{11}$ s. The orbital velocity is 9 $\times 10^{4}$ cm s$^{-1}$ which is in good agreement with the expected velocity for a circular orbit. The masses of the fragments at this stage are 8 $\times 10^{31}$ g each and the mass of the central density maximum is $10^{32}$ g. The total mass in the region not in fragments is $\approx 10^{32}$ g. The fragment mass is comparable to the distributed mass in the region and is still small compared with the total mass of the cloud; therefore it is too early to draw conclusions with regard to the final properties of the binary, as further accretion and fragmentation could occur. The later evolution of this configuration is discussed in §3.4.

3.2 Initial binary formation: high resolution

In order to test the effect of the grid resolution on the fragmentation process up to the time of 7.97 $\times 10^{11}$ s (Fig. 2f), we repeated the calculation with much higher spatial resolution. The finest subgrid now has its outer edge at 1.56 $\times 10^{15}$ cm and a resolution of 2.4 $\times 10^{13}$ cm. The initial collapse into a disk-like structure is very similar to the lower-resolution result. Fig. 3a shows the still almost homogeneous central region of this disk at a time of 6.63 $\times 10^{11}$ s. Shortly after that (Fig. 3b) the $m = 2$ perturbation grows into a central bar, the central part of which quickly wraps up into an inner disk with a central density maximum (Fig. 3c). The flow is directed into the disk through spiral arms which represent the original outer parts of the central bar. Note that the region outside the central disk is also disk-like, but with a larger scale height (Fig. 4). The scale height of the inner disk is probably smaller than the resolution limit of $2.4 \times 10^{13}$ cm. The outer disk has a well resolved thickness of $4 \times 10^{14}$ cm. Gas from the surroundings falls supersonically onto the outer disk through an accretion shock, where it is decelerated to subsonic velocities and then moves slowly toward the midplane and inwards toward the center. Fig. 5 shows the density distribution along the $x$-axis for $y = 0$ and $z = 0$. Coming inwards from the outer disk, which has $\log \rho \approx -13.8$, we reach the inner disk at a radius of $3 \times 10^{14}$ cm. The density in the inner disk is two orders of magnitude higher than in the outer disk. In the very center, the density rises again by more than one order of magnitude. One can also note that two density maxima are forming at the location where the material from the arms enters the disk (Fig. 3c). They are also seen as secondary density maxima in Fig. 5. As the arms wrap around the disk (Fig. 3d) the high-density points where the gas enters the disk rotate.

Up to this point in the evolution, the formation of an inner disk structure with spiral arms is very similar to that in the lower-resolution run, however the structure develops earlier and on a smaller scale. Because of this, the amount of mass in the inner disk region is smaller than before, and as a result a pronounced self-gravitating ring cannot form. As in the lower-resolution case, the axisymmetric disk structure is again unstable to bar formation, and it evolves into an inner spiral pattern which connects to the outer spiral arms of the original $m = 2$ perturbation (Fig. 3e). In the lower resolution run, a similar transition occurred (Fig. 2d) but in that case the spiral was dominated by the fragments which had formed earlier. In the higher resolution case there simply is not enough mass for fragmentation to occur at the stage of the formation of the inner disk. However, about half an orbital period ($2 \times 10^{10}$ sec.) later the disk has become massive enough to become bar-unstable, and it develops a spiral pattern with two condensations (Fig. 3f) with an initial separation of $10^{15}$ cm. The flow of gas inward along the outer spiral arms feeds the binary, whereas the gas interior to the binary accumulates in and around the center. In the lower-resolution case, the outer two of the four fragments grew in mass through accretion from the outer arms, whereas the inner fragments merged into the center, leading to a similar configuration: a binary on a circular orbit around a central density maximum. However note that the scale of the structure in Fig. 3f is a factor 4 smaller than that in Fig. 2f (which is the same factor as the increase in spatial resolution), and the time is considerably earlier.

Figs. 6a to 6f show the evolution of the system after the formation of the binary, up to a time of $7.95 \times 10^{11}$ s. First of all, note that the binary is stable on a circular orbit around a central density maximum, with a period of about $3 \times 10^{10}$ s and an orbital separation of $5 \times 10^{14}$ cm . During its first half orbit the external spiral arm which connects to the fragments winds up, creating a high-density circumbinary disk (Figs. 6a and 6b). One might expect, from the results of Bonnell & Bate (1994) that the binary can trigger further fragmentation in the disk as it grows in mass. In fact the onset of such a fragmentation is evident in Fig. 6b. An outer binary forms with an initial separation about twice that of the inner binary. As it rotates about the inner system (Figs. 6c to 6f) on a nearly circular orbit it accretes material with high angular momentum, increasing its mass and its separation.

At a time of $7.54 \times 10^{11}$ s (Fig. 6d) the infalling gas has continued to accumulate into a high-density disk which now also encloses the outer binary. An enlargement of Fig. 6d, showing the details of the circumbinary disk, circumstellar disks around the outer binary, the connecting bar, and the inner triple system, is shown in Fig. 7. At the end of the high-resolution calculation the inner binary has completed almost 3 orbits and its fragment masses are $4 \times 10^{31}$ g each. The mass associated with the central peak is $3 \times 10^{31}$ g. The total mass in the central region is therefore about the same as that found in the low-resolution calculation. There, however, the inner fragments merged into the center, probably because of insufficient resolution. The outer fragments of the high-resolution run have now reached an orbital separation of $3.5 \times 10^{15}$ cm and masses of about $6 \times 10^{31}$ g each. This outer binary has essentially the same properties as the surviving binary in the low-resolution case at the same time. Fig. 8 shows the system on a larger scale at the end of the calculation. Note that the outer binary has induced further fragmentation in the circumbinary disk, leading to a third binary with an initial separation of about $6 \times 10^{15}$ cm and an initial 90° phase shift with respect to the former outer binary.

3.3 Stability of the multiple system

The high resolution run indicates that the inner triple system is stable over at least four orbital periods. This result is a consequence of the fact that we have enforced point symmetry with respect to the center. In general one would expect that such a system would be unstable. Therefore we have made an approximate evaluation of the further evolution of the system of fragments by performing an $n$-body simulation for all five density maxima shown in Fig. 6f, adopting their positions, velocities and masses at $t = 7.908 \times 10^{11}$ s. The evolution of the inner triple system is probably well described by a pure $n$-body simulation as these fragments dominate the mass in the inner region. The outer fragments, however, are still accreting material from the surroundings. In addition, for such separations, the gravitational forces from the surrounding gas cloud cannot be completely neglected. We therefore concentrate mainly on the evolution of the inner triple system, but we include the forces on it caused by the outer binary. The integration of the orbits was done with a leapfrog method; the total energy of the system was conserved to within 1 part in $10^4$ over a time scale of $6 \times 10^{11}$ sec.

Fig. 9a shows the orbits of the three inner fragments over a time of $10^{11}$ s of $n$-body evolution. The later evolution of the system is shown in Fig. 9b. Figs. 10a and 10b show the time evolution of the orbital separation of the central object (called fragment 1) and one of the inner binary components (fragment 2) and of the orbital separation of the binary components (fragments 2 and 3), respectively, over a time of $5 \times 10^{11}$ s. The initial binary is stable for approximately $7 \times 10^{10}$ s with a mean $\log r_{1,2} = 14.45$ and a mean $\log r_{2,3} = 14.75$. Then the separation $r_{1,2}$ decreases as the central object spirals outwards, demonstrating, as expected, that the system is not stable (Fig. 9a). The central object picks up fragment 2 and goes into a close binary orbit with it, with a new orbital separation of $\log r_{1,2} = 14.18$. After a short transition period, the system settles into a new stable configuration where fragment 3 is now on an elliptical orbit about the binary with a period of about $10^{11}$ sec (Fig. 9b and Fig. 10b).

Continuing the $n$-body calculation up to $10^{12}$ sec into the evolution, we find that the outer binary, which in this idealized situation is on an elliptical orbit, eventually interacts with the inner triple system making it unstable again. The end result of this interaction is uncertain, but possible outcomes are merger of the inner and outer binaries, formation of a hierarchical binary system, or ejection of individual fragments through a close encounter with more massive components. Note however that because of the finite resolution of the hydro code, close encounters will most likely end in a merger.

3.4 Later hydrodynamical evolution

We have reached the same point in time (Fig. 6f) with the high resolution hydrodynamical calculation as we reached with moderate resolution (Fig. 2f), and the results are indeed very similar: an outer binary with a total mass of about $1.5 \times 10^{32}$ g orbits around a high-density central peak with a total mass of $10^{32}$ g. In the high resolution case the central peak is resolved into a triple system, the later evolution of which was discussed in section 3.3. We now concentrate on the evolution of the outer binary and the surrounding disklike gas cloud, which still contains most of the mass. It is important to investigate whether further fragmentation can occur on larger scales, as Fig. 8 suggests. Because the high -resolution run consumes a great deal of computer time and because grid boundary effects result in the accumulation of numerical errors, we now return to moderate resolution and continue the run illustrated in Fig. 2.

As the evolution progresses, the binary components accrete more mass and angular momentum. At $t = 8.54 \times 10^{11}$ sec (Fig. 11a) the former central high-density disk has become bar unstable and has transformed into a spiral pattern. The central object becomes tidally elongated and begins to lose its gas to the binary through the connecting bar. At $t = 1.07 \times 10^{12}$ s the $m=2$ potential perturbation of the binary creates, in the surrounding differentially rotating lower-density disk, a trailing spiral pattern (Fig. 11b). Over the next $2 \times 10^{11}$ sec (about half an orbital period) the spiral winds up which results in a new circumbinary disk. As in the high-resolution case, the binary induces formation of a new pair of outer fragments with a separation about twice that of the original pair and with a 90° phase shift (Fig. 11c; cf. Fig. 8).

It seems as though the outer density maxima might condense into another binary; however as Fig. 11d demonstrates, the outer knots are pulled inwards by the inner binary, become tidally elongated, and finally merge at $t = 1.36 \times 10^{12}$ sec. This merger results in an increase of the mass and angular momentum of the binary, and in an orbit with higher eccentricity and larger separation (Fig. 11e). Figs. 11e and 11f show the late evolution of the binary over an additional full orbital period, which now is about $8 \times 10^{11}$ sec. The final separation is about $2 \times 10^{16}$ cm, similar to what one would expect if a substantial fraction of the mass and angular momentum of the cloud had condensed into a binary, which turns out to be the case. A further stage of circumbinary disk formation with subsequent fragmentation farther out is not possible because of the low density of the outer gas. However, note the well-marked circumstellar disks around the individual binary components, which contain a substantial fraction of the uncondensed gas mass. Also, a connecting bar forms between the components as a result of the binary's gravitational effect on the gas. Fig. 11e shows that the bar has fragmented; however the resolution is not sufficient to follow this process in detail. The fact that a massive binary could induce the formation of a fragmenting bar out of a disk has already been demonstrated in an earlier paper (BB).

At the end of the calculation the fragments of the bar have fallen into the center and have merged, forming a central density maximum with a surrounding disk. The central mass is $4.2 \times 10^{32}$ g, similar to the masses of the binary components, which are each $4.7 \times 10^{32}$ g. About 70% of the mass of the initial cloud mass has now condensed into fragments. The specific angular momentum of the orbit is $4 \times 10^{20}$ cm$^2$ s$^{-1}$. The specific spin angular momentum of each fragment is estimated to be $4 \times 10^{19}$ cm$^2$s$^{-1}$, a factor of about 15 less than that of the original cloud ($6.0 \times 10^{20}$ cm$^2$s$^{-1}$). The parameters $\alpha$ and $\beta$ for the fragments are estimated to be, respectively, $2.2 \times 10^{-2}$ and 0.115. It is of interest to determine whether the fragments could subfragment during the subsequent collapse through the adiabatic phase. An estimate can be made by applying the criterion for fragmentation for an adiabatic collapse with $\gamma = 1.4$ (Hachisu et al 1987, Tohline 1981, Boss 1981, Miyama 1992). For a uniform initial state fragmentation occurs if $\alpha < 0.09\beta^{0.2}$. Thus in our case subfragmentation and the formation of a hierarchical system is indicated; however additional detailed calculations are required to determine the properties of the resulting system. By way of contrast, the fragments obtained by Boss (1991) in case C4 and by Boss (1993a) are not likely to subfragment according to this criterion.

## 4 CONCLUSIONS

Several conclusions can be reached from the results of these calculations. First, even after some fragments have appeared in a calculation, the system is likely to evolve further. Therefore no conclusion regarding the fragmentation of a system should be drawn from calculations which do not follow the dynamical evolution of the initial fragments for at least one orbital period. Second, at the time of initial fragmentation, most of the original mass and angular momentum of the cloud still lies outside the fragments. The further evolution will be influenced by accretion onto existing fragments, mergers, as well as by possible further fragmentation induced by the gravitational interaction of fragments with the surrounding disk. Such important phases of secular evolution would be missed if the calculation were stopped too early.

For the particular initial conditions that we chose, we find an interesting pattern of how binaries are created sequentially by the action of pre-existing binaries with their circumbinary disks. However, binaries formed in this way have initial orbital separations which are only about twice as large as those of the binaries that generated them. The resulting configuration is not stable, leading to secular merging or to redistribution of fragments into more stable hierarchical binary configurations. We have actually observed both effects in our calculations. First, according to an $n$-body calculation, the inner triple system evolves to a hierarchical system. Second, according to the hydrodynamical calculations a set of outer fragments, induced in a circumbinary disk, later merges with the binary that induced them. Both kinds of calculations have their limitations: the $n$-body simulations neglect the effect of accretion, circumstellar disks, and the overall gravitational field of the cloud, while the hydrodynamical calculation, because of finite resolution, tends to produce mergers, suppressing the formation of hierarchical systems or escapees.

The net result of the calculation seems to be independent of a significant change in numerical resolution. A high-resolution calculation shows a very similar formation scenario, but on smaller scales, than does a low-resolution calculation. The sequence of events in fragment formation and coalescence may be different for different resolutions, but after the initial transients have passed, the end result is about the same. In particular, the mechanism by which binary formation is triggered in a circumbinary disk (by the inner binary) does not depend on the numerical resolution. Grid boundaries in a nested grid scheme could conceivably lead to serious numerical effects, resulting in artificial fragmentation in a high-density region. We do not believe that our binaries are a result of such an effect; in the cases of induced fragmentation shown in Figs. 6b and 11c the new fragments form far from any grid boundary, while in the example shown in Fig. 8 the new fragments form in inflowing gas several zones outside a grid boundary.

For these particular input parameters, which involve a fairly high angular momentum with respect to those observed in interstellar cloud cores, the end product is a binary, containing a major fraction of the mass and angular momentum. Its interaction with the triple system that was formed in the central regions in the high-resolution case could result in additional mergers, or in new stable orbits with the small fragments being in close orbit around one of the major binary components, or even in an ejection of one or more low-mass fragments.

These results show that a relatively simple initial condition leads to the formation of a binary which closely reflects the initial conditions of total mass and total angular momentum of the cloud core. If this is what typically happens in a cloud collapse, then the distribution of angular momenta and of orbital separations of massive, long-period binaries would simply be a consequence of the initial formation process of cloud cores rather than a consequence of fragmentation theory. On the other hand, short-period binaries with at least one low-mass component, as well as low-mass single stars, seem to form quite naturally as byproducts during the process of formation of the major binary. The finer details of fragmentation theory, the first indications of which are noticeable in our high-resolution run, might then be very important. The underlying goal of these calculations is to explain the distribution of binary periods and the IMF. To make further progress toward this goal we will need detailed high-resolution runs, such as that presented here, for a large set of values of $\alpha$ and $\beta$, selecting especially more realistic (lower) values of $\beta$. Also, the assumed initial density pertubation very likely affects the outcome of the calculation. More general types of perturbations have to be studied, for example, a superposition of various modes. The technique of nested grids provides enough resolution for the study of the details of local fragmentation. The results of future parameter studies along these lines will be reported in future papers.

**ACKNOWLEDGMENTS**

This work was supported in part through National Science Foundation grant AST-9315578 and in part through a special NASA astrophysics theory program which supports a joint Center for Star Formation Studies at NASA/Ames Research Center, UC Berkeley, and UC Santa Cruz. AB thanks the staff of UCO/Lick Observatory for the hospitality during his visit, and PB thanks the staff of the Max Planck Institute for Astrophysics in Garching for the hospitality during his visit.

## FIGURE CAPTIONS

**Figure 1.** Moderate resolution case at a time of $7.49 \times 10^{11}$ s. Contours of equal density (in g cm$^{-3}$) on the $(x,y)$ plane at $z=0$ are plotted on the innermost grid with a maximum density of log $\rho_{max}$ =−11.39 and a contour interval $\Delta \log \rho = 0.159$. Arrows indicate velocity vectors with length proportional to speed and with maximum vector length corresponding to $v_{max}$. The spatial scale is given in cm, and $v_{max} = 1.68 \times 10^5$ cm s$^{-1}$.

**Figure 2.** Evolution of the moderate-resolution case. Symbols and curves have the same meaning as in Fig. 1. The horizontal scale in each frame is the same as the vertical scale. (a; upper left) $t = 7.543 \times 10^{11}$ s; log $\rho_{max}$ = -11.37; $\Delta \log \rho = 0.244$; $v_{max} = 1.70 \times 10^5$ cm s$^{-1}$. (b; upper right) $t = 7.631 \times 10^{11}$ s; log $\rho_{max}$ = -11.32; $\Delta \log \rho = 0.246$; $v_{max} = 1.82 \times 10^5$ cm s$^{-1}$. (c; center left) $t = 7.682 \times 10^{11}$ s; log $\rho_{max}$ = -11.29; $\Delta \log \rho = 0.248$; $v_{max} = 1.96 \times 10^5$ cm s$^{-1}$. (d; center right) $t = 7.748 \times 10^{11}$ s; log $\rho_{max}$ = -11.08; $\Delta \log \rho = 0.259$; $v_{max} = 3.13 \times 10^5$ cm s$^{-1}$. (e; lower left) $t = 7.787 \times 10^{11}$ s; log $\rho_{max}$ = -10.96; $\Delta \log \rho = 0.265$; $v_{max} = 3.78 \times 10^5$ cm s$^{-1}$. (f; lower right) $t = 7.972 \times 10^{11}$ s; log $\rho_{max}$ = -10.95; $\Delta \log \rho = 0.265$; $v_{max} = 2.24 \times 10^5$ cm s$^{-1}$.

**Figure 3.** Evolution of the high-resolution case. Symbols and curves have the same meaning as in Fig. 1, except that the spatial scale is reduced by a factor 4. The arrangement of frames is the same as in Fig. 2. (a) $t = 6.628 \times 10^{11}$ s; log $\rho_{max}$ = -14.24; $\Delta$ log $\rho$ = 0.024; $v_{max} = 1.81 \times 10^4$ cm s$^{-1}$. (b) $t = 6.863 \times 10^{11}$ s; log $\rho_{max}$ = -13.10; $\Delta$ log $\rho$ = 0.018; $v_{max} = 4.49 \times 10^4$ cm s$^{-1}$. (c) $t = 7.086 \times 10^{11}$ s; log $\rho_{max}$ = -9.991; $\Delta$ log $\rho$ = 0.316; $v_{max} = 1.45 \times 10^5$ cm s$^{-1}$. (d) $t = 7.135 \times 10^{11}$ s; log $\rho_{max}$ = -9.955; $\Delta$ log $\rho$ = 0.318; $v_{max} = 1.58 \times 10^5$ cm s$^{-1}$. (e) $t = 7.190 \times 10^{11}$ s; log $\rho_{max}$ =-9.922; $\Delta$ log $\rho$ = 0.316; $v_{max} = 1.66 \times 10^5$ cm s$^{-1}$. (f) $t = 7.197 \times 10^{11}$ s; log $\rho_{max}$ = -9.836; $\Delta$ log $\rho$ = 0.324; $v_{max} = 1.83 \times 10^5$ cm s$^{-1}$.

**Figure 4.** The vertical ($x, z$ plane at $y = 0$) structure of the high-resolution case at $t = 7.086 \times 10^{11}$ s. Contours of equal density and velocity vectors are shown, as in Fig. 1. log $\rho_{max}$ = -9.993; $\Delta$ log $\rho$ = 0.242; $v_{max} = 1.75 \times 10^5$ cm s$^{-1}$.

**Figure 5.** Density as a function of $x$ at $y = 0$, $z = 0$ for the high-resolution case at $t = 7.086 \times 10^{11}$ s.

**Figure 6.** Evolution of the high-resolution case. The scale is a factor 2 larger than that in Fig. 3. The entire innermost grid is shown. Symbols and curves have the same meaning as in Fig. 1. The arrangement of frames is the same as in Fig. 2. (a) $t = 7.226 \times 10^{11}$ s; log $\rho_{max}$ = -9.70; $\Delta$ log $\rho$ = 0.332; $v_{max} = 1.81 \times 10^5$ cm s$^{-1}$. (b) $t = 7.259 \times 10^{11}$ s; log $\rho_{max}$ = -9.70; $\Delta$ log $\rho$ = 0.332; $v_{max} = 2.27 \times 10^5$ cm s$^{-1}$. (c) $t = 7.365 \times 10^{11}$ s; log $\rho_{max}$ = -9.77; $\Delta$ log $\rho$ = 0.328; $v_{max} = 3.39 \times 10^5$ cm s$^{-1}$. (d) $t = 7.543 \times 10^{11}$ s; log $\rho_{max}$ = -9.55; $\Delta$ log $\rho$ = 0.340; $v_{max} = 4.24 \times 10^5$ cm s$^{-1}$. (e) $t = 7.747 \times 10^{11}$ s; log $\rho_{max}$ = -9.43; $\Delta$ log $\rho$ = 0.346; $v_{max} = 4.88 \times 10^5$ cm s$^{-1}$. (f) $t = 7.947 \times 10^{11}$ s; log $\rho_{max}$ = -9.53; $\Delta$ log $\rho$ = 0.340; $v_{max} = 3.78 \times 10^5$ cm s$^{-1}$.

**Figure 7.** Enlargement of Fig. 6d.

**Figure 8.** The end of the high-resolution run at a time of $8.03 \times 10^{11}$ s. Symbols and curves have the same meaning as in Fig. 1. Log $\rho_{max} = -10.94$; $\Delta$ log $\rho = 0.174$; $v_{max} = 2.55 \times 10^5$ cm s$^{-1}$. The linear scale is a factor 4 larger than that in Fig. 6.

**Figure 9.** Orbits of the three inner fragments of the high-resolution run starting at $t = 7.908 \times 10^{11}$ sec. Fig. 9a shows the evolution of these three objects for the first $10^{11}$ sec of the $n$-body calculation. Solid lines: fragments 1 and 3; dashed line: fragment 2. Fig. 9b shows the evolution from $t = 1.6 \times 10^{11}$ sec to $t = 5.5 \times 10^{11}$ sec, measured from the beginning of the $n$-body calculation. The solid curve corresponds to the center of mass of the binary consisting of fragments 1 and 2; the dashed line shows the orbit of fragment 3.

**Figure 10.** The orbital separation as a function of time of fragments 1 and 2 (Fig. 10a; left) and fragments 2 and 3 (Fig. 10b; right) during the $n$-body calculation.

**Figure 11.** Late evolution of the moderate-resolution case. Symbols and curves have the same meaning as in Fig. 1. The arrangement of frames is the same as in Fig. 2. (a) $t = 8.539 \times 10^{11}$ s; log $\rho_{max}$ = -11.05; $\Delta$ log $\rho$ = 0.260; $v_{max} = 3.73 \times 10^5$ cm s$^{-1}$. (b) $t = 1.067 \times 10^{12}$ s; log $\rho_{max}$ = -11.05; $\Delta$ log $\rho$ = 0.260; $v_{max} = 4.22 \times 10^5$ cm s$^{-1}$. (c) $t = 1.248 \times 10^{12}$ s; log $\rho_{max}$ = -11.54; $\Delta$ log $\rho$ = 0.235; $v_{max} = 3.31 \times 10^5$ cm s$^{-1}$. (d) $t = 1.343 \times 10^{12}$ s; log $\rho_{max}$ = -11.90; $\Delta$ log $\rho$ = 0.216; $v_{max} = 3.55 \times 10^5$ cm s$^{-1}$. (e)

$t = 1.520 \times 10^{12}$ s; log $\rho_{max}$ = -11.84; $\Delta$ log $\rho$ = 0.219; $v_{max} = 2.97 \times 10^5$ cm s$^{-1}$. (f)
$t = 2.202 \times 10^{12}$ s; log $\rho_{max}$ = -11.96; $\Delta$ log $\rho$ = 0.213; $v_{max} = 2.82 \times 10^5$ cm s$^{-1}$.